\documentclass[12pt,reqno,a4paper]{amsart}

\usepackage{lipsum,subeqnarray}
\usepackage{amsfonts}
\usepackage{graphicx}
\usepackage{amsmath}
\usepackage{mathtools}
\usepackage{amssymb}
\usepackage{enumerate}
\usepackage{slashed}
\usepackage{graphicx}
\usepackage{newlfont}
\usepackage{amsrefs}
\usepackage{comment}
\usepackage[normalem]{ulem}
\usepackage{mathtools}
\usepackage{accents}
\usepackage{leftidx}
\usepackage{bbm}
\usepackage{amsthm}
\usepackage{amsfonts}

\numberwithin{equation}{section}

\theoremstyle{plain}
\newtheorem{theorem}{Theorem}[section]

\newtheorem{proposition}[theorem]{Proposition}
\newtheorem*{theorem*}{Conjecture}

\newtheorem*{Thm}{Theorem}

\usepackage{xcolor}

\theoremstyle{definition}

\theoremstyle{remark}
\newtheorem{remark}{Remark}
\usepackage{stackengine}
\usepackage{subeqnarray} 

\stackMath
\newcommand\tenq[2][1]{%
 \def\useanchorwidth{T}%
  \ifnum#1>1%
    \stackunder[0pt]{\tenq[\numexpr#1-1\relax]{#2}}{\scriptscriptstyle\sim}%
  \else%
    \stackunder[1pt]{#2}{\scriptscriptstyle\sim}%
  \fi%
}

\begin{document}
\title[Counter examples to the PI and PMT assuming WEC]{Spherically symmetric counter examples to the Penrose inequality and the positive mass theorem under the assumption of the weak energy condition}
 
\author{Jaroslaw S. Jaracz}

\begin{abstract}
Of the various energy conditions which can be assumed when studying mathematical general relativity, intuitively the simplest is the weak energy condition $\mu\geq 0$ which simply states that the observed mass-energy density must be non-negative. This energy condition has not received as much attention as the so-called dominant energy condition. When the natural question of the Penrose inequality in the context of the weak energy condition arose, we could not find any results in the literature, and it was not immediately clear whether the inequality would hold, even in spherical symmetry. This led us to constructing a spherically symmetric asymptotically flat initial data set satisfying the weak energy condition which violates the "usual" formulations of the Penrose conjecture. We remark this does not contradict the result in \cite{MalecMurchada} since there the authors assume, in addition to the weak energy condition, that the initial data is maximal, which implies positive scalar curvature. Our construction itself is quite elementary. However, the consequences of the counterexample are quite interesting. The Penrose inequality was conjectured by Penrose in \cite{Penrose} using certain heuristic arguments which, as we discuss, continue to hold in the case of the weak energy condition. Yet, a counter example exists. Moreover, the methods developed naturally led to the construction of a counter example to the positive mass theorem assuming the weak energy condition. The counter example can be constructed to be diffeomorphic to $\mathbb{R}^3$ and to contain no minimal surfaces and no future apparent horizons.
\end{abstract}
\maketitle

\section{Introduction and Main Theorems}\label{SEC:Intro}

\subsection{The Penrose Conjecture}

The Penrose inequality has been one of the most famous open conjectures in mathematical general relativity. Conjectured by Roger Penrose in the 1970's using a heuristic argument based on the established view point of gravitational collapse \cite{Penrose}, it relates the total mass $m$ of a spacetime to the surface area $A$ of a black hole in the spacetime via the inequality
\begin{equation}
    m \geq \sqrt{\frac{A}{16 \pi }}= \frac{1}{2}\rho
\end{equation}
where we define the area radius $\rho$ by $A=4\pi \rho^2$. We will discuss the heuristic argument in detail later on.

It turns out that the notion of mass in general relativity is a tricky concept. In fact, there is no accepted notion of quasi-local mass which has all the desired and expected properties. As a result one needs a mathematically precise notion of mass. This turns out to be given by the ADM formalism, where the ADM energy can be thought of as the total mass of the system from the point of view of an observer at infinity. The ADM energy is only well defined for certain types of coordinate systems, which are referred to as asymptotically flat. We mention that there is some inconsistency for the terminology used by different authors relating to "ADM energy" and "ADM mass." We will discuss this inconsistency in more detail once we give the relevant definitions.

A special case, known as the Riemannian Penrose inequality, was proven in the late 1990's for an asymptotically flat initial data set $(M,g)$ by Huisken and Illmanen using a weak version of the inverse mean curvature flow \cite{HuiskenIlmanen}, and independently by Hubert Bray using a conformal flow of metrics \cite{Bray}. In these cases, the black hole is represented by a minimal surface and the initial data set must have non-negative scalar curvature $R\geq 0$ (or to satisfy some assumptions which imply this condition). The $m$ in these cases is given by the ADM energy.

The Penrose inequality for a general asymptotically flat initial data set $(M, g, k)$ where $k$ is the extrinsic curvature remains an open problem. It has been proven in the case of spherical symmetry where $m$ is given by the ADM energy, assuming the so-called dominant energy condition where the black hole is mathematically represented by an outermost future or past apparent horizon \cite{Hayward}. Hence, a popular form of the Penrose conjecture is the following:

\begin{theorem*}[Penrose Inequality, ADM Energy Version] 
Let $(M, g, k)$ be an asymptotically flat initial data set satisfying appropriate fall-off conditions and the dominant energy condition, with boundary $\partial M$ consisting of an outermost apparent horizon. Let $A=A_{min}(\partial M)$ denote the outermost minimal area enclosure of $\partial M$. Then
\begin{equation*}
    E_{ADM}\geq \sqrt{\frac{A}{16\pi}}
\end{equation*}
where $E_{ADM}$ is the ADM energy.
\end{theorem*}

Another formulation of the conjecture replaces the ADM energy by the ADM mass. 

\begin{theorem*}[Penrose Inequality, ADM Mass Version] 
Let $(M, g, k)$ be an asymptotically flat initial data set satisfying appropriate fall-off conditions and the dominant energy condition, with boundary $\partial M$ consisting of an outermost apparent horizon. Let $A=A_{min}(\partial M)$ denote the outermost minimal area enclosure of $\partial M$. Then
\begin{equation*}
    m_{ADM}\geq \sqrt{\frac{A}{16\pi}}
\end{equation*}
where $m_{ADM}$ is the ADM mass.
\end{theorem*}

It is then natural to ask if the dominant energy condition in the above conjecture could be replaced by the weak energy condition (for the definitions of all the relevant quantities, see Section \ref{Definitions}). The answer is negative, and is encapsulated in the following theorem.

\begin{theorem} \label{MainTheorem}
There exists an asymptotically flat initial data set $(M, g, k)$ with $\partial M$ consisting of an outermost future apparent horizon, satisfying the usual fall-off conditions
\begin{align} \label{FallOffConditions}
    | D^{\lambda} ( g_{ij} - \delta_{ij}) | \leq Cr^{-1-|\lambda|}, \quad |R|\leq Cr^{-3}, \quad |k|\leq Cr^{-2}, \quad |Tr_g k | \leq Cr^{-2} 
\end{align}
for some constant $C$, the weak energy condition, and for which 
\begin{equation} \label{MainTheoremInequality1}
    \sqrt{  \frac{A}{16\pi} } > E_{ADM}
\end{equation}
and
\begin{equation} \label{MainTheoremInequality2}
     \sqrt{  \frac{A}{16\pi} } > m_{ADM} 
\end{equation}
where $A=A_{min}(\partial M)$ is the area of the outermost minimal area enclosure of $\partial M$. Here $E_{ADM}$ is the ADM energy and $m_{ADM}$ is the ADM mass.
\end{theorem}

Once again, we mention this does not contradict the result in \cite{MalecMurchada} as in that paper, in addition to the weak energy condition, the authors assume that the initial data set is maximal ($Tr_g k=0$) which combined with \eqref{ConstraintEquations} and $\mu\geq 0$ implies that $R\geq 0$. Hence, the combination of the weak energy condition with the maximality assumption is quite a strong condition on the initial data set.

The initial data set we construct will be spherically symmetric and in fact both $R$ and $k$ will be compactly supported. In Section \ref{Definitions} we give precise definitions of all relevant quantities and collect formulas for spherically symmetric metrics. Then in Section \ref{Proof} we use these formulas to construct a spherically symmetric metric with the desired properties. Before that however, we discuss Penrose's heuristic argument and discuss the significance of our theorem.

\subsection{Comments on Penrose's Heuristic Argument}

The heuristic argument for the inequality was first given by Penrose in \cite{Penrose} and it depends on several ingredients. We only need to focus on the ingredient which depends on the energy condition. An excellent exposition is given in the introduction of the now classical review article \cite{Mars_2009}, and that is where we refer the reader interested in the details of the other ingredients.

The key ingredient we are interested in is the black hole area law. For the area law to hold something called the null energy condition (defined later) must hold. However, the weak energy condition implies the null energy condition. Thus, if the spacetime satisfies the weak energy condition, the area law holds, and combining this with the other usual ingredients leads to the Penrose inequality. 

For completeness we give a sketch of the idea. One looks at a spacetime $(\mathcal{M}, \mathfrak{g})$ which is strongly asymptotically predictable, admits a complete future null infinity, and contains an apparent horizon $\Sigma$. One then takes a spacelike asymptotically flat slice with ADM energy $E_{ADM}$, whose intersection with $\Sigma$ is some surface $S_i$ with area $|S_i|$ (the $i$ standing for initial). Now, taking any slice of $\Sigma$ in the causal future of $S_i$, which we denote by $S_f$, the black hole area theorem states $|S_f|\geq |S_i|$. Then one makes the physical assumptions that the spacetime will eventually settle down to some equilibrium configuration and that all of the matter fields will eventually be swallowed up by the black hole. These assumptions imply that the spacetime must settle down to a Kerr blackhole, which satisfies $A_K \leq 16\pi m_K^2$ where $A_K$ is the area of the event horizon (which turns out to be independent of the slice of the Kerr spacetime) and $m_K$ is the Kerr mass parameter. Moreover, $m_K$ should be asymptotic to the Bondi energy. Since gravitational waves carry positive energy the Bondi energy must be nonincreasing to the future. Then, if one assumes the Bondi energy approaches the ADM energy of the initial slice (which requires certain additional assumptions) one obtains 
\begin{equation*}
    \sqrt{\frac{|S_i|}{16\pi}} \leq \sqrt{\frac{|S_f|}{16\pi}} \leq  \sqrt{\frac{|A_K|}{16\pi}} \leq m_K \leq E_{B_f} \leq E_{B_i} = E_{ADM}
\end{equation*}
where $E_{B_i}$ and $E_{B_f}$ are the initial and future Bondi energy. 

As we see, as long as we have the black hole area theorem, which we do, this heuristic argument goes through fine. And yet, we have a counterexample. This is quite mysterious and thus interesting.

\subsection{The Positive Mass Theorem}
The positive mass theorem states that for any spacetime the mass is positive, $m\geq 0$. Once again, the precise mathematical formulation of this statement is given in terms of asymptotically flat initial data sets, an appropriate energy condition, and the ADM energy. The following theorem was famously proven by Schoen and Yau. 

\begin{Thm}[Schoen \& Yau, 1981 \cite{SchoenYau2}]
Let $M$ be an asymptotically flat $3$-dimensional manifold without boundary satisfying the dominant energy condition, with finitely many ends $M_k$ each satisfying the fall-off conditions
\begin{equation} \label{FallOffConditionsSY} 
     | D^{\lambda} ( g_{ij} - \delta_{ij}) | \leq Cr^{-1-|\lambda|}, \quad |D^{\lambda}R|\leq Cr^{-4-|\lambda|}, \quad |D^{\lambda}k_{ij}|\leq Cr^{-2-|\lambda|}, \quad |Tr_g k | \leq Cr^{-3} 
\end{equation}
in the asymptotically flat coordinates, with $\lambda$ being any multiindex with $|\lambda|\leq 2$. Then 
\begin{equation*}
    E_k \geq 0
\end{equation*}
where $E_k$ is the ADM energy of the $k$-th end.
\end{Thm}

We remark that in \cite{SchoenYau2}, the authors use the term "ADM mass" for what we call the "ADM energy" and so with our terminology we should refer to it as the positive energy theorem. However, as is customary we continue to use the term "positive mass theorem" when referring to it.

As a byproduct of our proof of Theorem \eqref{MainTheorem}, we actually construct an asymptotically flat manifold with negative ADM energy. However, that initial data set does have a boundary which happens to be an apparent horizon. However, we'd like our data set to satisfy all they hypotheses of the above theorem, with the exception of satisfying the weak energy condition as opposed to the dominant energy condition. Fortunately, with slight modifications to the proof of Theorem \ref{MainTheorem}, we can make $M \cong \mathbb{R}^3$ and moreover be free of minimal surfaces and future apparent horizons.

\begin{theorem}\label{MainTheorem2}
There exists an asymptotically flat initial data set $(M, g, k)$ with $M$ a $3$-dimensional manifold without boundary with a single end satisfying the fall-off conditions \eqref{FallOffConditionsSY} and satisfying the weak energy condition such that 
\begin{equation*}
    E_{ADM}<0.
\end{equation*}
Moreover, we have $M\cong \mathbb{R}^3$ and the initial data set doesn't contain any minimal surfaces or future apparent horizons.
\end{theorem}

\section{Definitions and Standard Formulas}\label{Definitions}

\subsection{Asymptotic Flatness and the ADM Formalism} We will consider an initial data set $(M, g, k)$ where $M$ is a $3$-manifold, $g$ a Riemannian metric, and $k$ is a symmetric $2$-tensor, the extrinsic curvature. The general discussion of asymptotically flat ends and the ADM mass can be found in \cite{ADM} and \cite{Bartnik}. Briefly, we say that $M$ is asymptotically flat, if for the complement of some compact set $K$ it is a union of finitely many ends $M_i$
\begin{equation*}
    M \smallsetminus K = \cup_{i=1}^n M_i
\end{equation*}
where each $M_i \cong \mathbb{R}^3 \smallsetminus B$ for some ball $B$, and on each end there exist coordinates such that $g$ and $k$ in these coodinates satisfy certain fall-off conditions. Different authors take different fall-off conditions, but for our purposes we take the fall-off conditions \eqref{FallOffConditions}, which are standard. There, $\delta$ is the Euclidean metric, $r=\sqrt{x^2+y^2+z^2}$ the standard Euclidean radius, $D^\lambda$ is a derivative operator with respect to the Euclidean coordinates, and $\lambda$ is a multi-index. We have 
\begin{equation*}
    |k|^2=k_{ij}k^{ij}, \quad Tr_g k = g^{ij}k_{ij}
\end{equation*}
as usual.  

For an asymptotically flat end, the ADM energy and ADM momentum are defined by 
\begin{align} \label{ADMEnergy} 
E_{ADM}=\lim_{r\rightarrow \infty}\frac{1}{16\pi} \sum_{i, j} \int_{S_r} (g_{ij,i}-g_{ii,j})\nu^j dS_r \\ \label{ADMMomentum}
P_i = \lim_{r\rightarrow \infty} \frac{1}{8\pi}  \sum_j \int_{S_r} \left(  k_{ji}\nu^j - (Tr_g k) \nu_i \right) dS_r
\end{align}
where $S_r$ are coordinate spheres of radius $r$ and $\nu^j$ is the outward unit normal \cite{ADM}. 
One then defines the ADM mass by
\begin{equation} \label{ADMMass}
    m_{ADM}=\sqrt{ E^2_{ADM} - |P|^2  }
\end{equation}
These expressions are coordinate dependent. However, it is well known \cite{Bartnik}, \cite{ChruscielADM} that with the appropriate fall-off conditions the above quantities with respect to a chosen end are geometric invariant and doesn't depend on the choice of asymptotically flat coordinates in the particular end.

Here we mention more thoroughly the inconsistency in terminology that sometimes occurs, namely some authors refer to the quantity defined by \eqref{ADMEnergy} as ADM energy and some as ADM mass. Just for some examples, \eqref{ADMEnergy} is referred to as the ADM energy in \cites{Mars_2009, Wald} and the ADM mass in \cites{ChruscielNotes, Bray, HuiskenIlmanen, SchoenYau2}. In the latter cases, the formula \eqref{ADMMass} is usually not discussed. We will follow the terminology used in \cites{Mars_2009, Wald}.

\subsection{Energy Conditions and Constraint Equations}

Consider a $3+1$ spacetime $(\mathcal{M}, \mathfrak{g})$ with stress tensor $T_{ab}$. The spacetime is said to satisfy the null energy condition if 
\begin{equation*}
   \nu= T_{ab}K^aK^b \geq 0
\end{equation*}
for any null vector $K$.
It is said to satisfy the weak energy condition if 
\begin{equation*}
   \mu= T_{ab}X^aX^b \geq 0
\end{equation*}
for any timelike vector $X$. The interpretation of $\mu$ is that it is the mass-energy density observed by an observer traveling with tangent vector $X$. The quantity $\nu$ can be thought of as a limit of $\mu$. The spacetime is said to satisfy the dominant energy condition if for all future directed timelike $X$ we have
\begin{equation*}
    -T^a_bX^b
\end{equation*}
is a future directed timelike or null vector. The interetation here is that the speed of mass-energy flow is always less than the speed of light. The dominant energy condition implies the weak energy condition, which implies the null energy condition \cite{Wald}. To see that the weak energy condition implies the null energy condition one can simply take a sequence of timelike vectors $X$ converging to the null vector $K$.

Taking the trace of the Gauss-Codazzi equations, one finds that an initial data set for the Einstein equations must satisfy the constraint equations
\begin{align} \label{ConstraintEquations}
\begin{split}
16\pi\mu=R+(Tr_gk)^2-|k|^2\\
8\pi J_i=\nabla^j (k_{ij}-(Tr_g k)g_{ij})
\end{split}
\end{align}
where $R$ is the scalar curvature, $\nabla^j$ denotes covariant differentiation, $\mu$ is again the mass-energy density, and $J_i$ the components of the momentum density. 
Apriori, there are no constraints on the matter or momentum densities. 

For an initial data set the weak energy condition takes the form
\begin{equation}\label{WEC}
    \mu \geq 0 
\end{equation}
while the dominant energy condition takes the form
\begin{equation*}
    \mu \geq |J|_g.
\end{equation*}
Based on the experimental behavior of classical matter, either one of these energy conditions is physically reasonable. We say classical matter, because the quantum mechanical Casimir effect produces energy densities which are negative relative to the vaccum energy \cite{Schwartz}.

Mathematically, being a stronger condition, the dominant energy condition allows the proof of stronger statements. For exmaple, the proof of the positive mass theorem \cite{SchoenYau2} requires the assumption of the dominant energy condition. Similarly, in \cite{Hayward} it was proven that in spherical symmetry for an outermost apparent horizon with area $A$ one has
\begin{equation*}
    E_{ADM}\geq \sqrt{\frac{A}{16\pi}}
\end{equation*}
which, assuming $k$ falls off sufficiently fast to guarantee $P=0$ and thus $m_{ADM}=E_{ADM}$, proves the Penrose inequality in spherical symmetry.

\subsection{Null expansions and apparent horizons}

Given a two dimensional surface $S$ inside $M$ we can calculate the future (+) and past (-) null expansion at each point of the surface, defined by
\begin{equation*}
    \theta_\pm = H_S \pm Tr_S k
\end{equation*}
where $H_S$ indicates the mean curvature of the surface, and $Tr_S k$ indicates the trace of $k$ restricted to $S$ calculated with respect to the induced metric. The null expansions measure the convergence and divergence of past and future directed null geodesics. A future or past apparent horizon is defined by  
\begin{equation*}
    \theta_{\pm}=H_S \pm Tr_S k = 0.
\end{equation*}
An apparent horizon is one way of modeling a black hole. The idea is that if one was to emit a pulse of light from the surface of the black hole, the resulting shell of light would not expand due to the strength of the gravitational field. That is precisely the behavior observed at an apparent horizon as measured by the null expansions. Hence, apparent horizons are a popular way of mathematically modelling black holes in initial data sets.

\subsection{The inverse mean curvature flow and the Hawking mass}

The inverse mean curvature flow (IMCF) is a flow of surfaces in the direction of the outward unit normal where the speed equals $1/H$ where $H$ is the mean curvature. In general, the IMCF is not smooth, but a weak formulation was developed by Huisken and Ilmanen for proving the Riemannian Penrose inequality \cite{HuiskenIlmanen}. In the case of spherical symmetry, the IMCF has a particularly simple form, which we will exploit.

The Hawking mass of a surface $S$ is defined by 
\begin{equation} \label{HawkingMass}
    m_H(S) = \sqrt{\frac{|S|}{16\pi}} \left( 1 - \frac{1}{16\pi} \int_S H^2 dS \right)
\end{equation}
where $H$ is the mean curvature of the surface and $|S|$ its area. In addition, if $N_t$ are the flow surfaces of the weak IMCF for $t\in[0, \infty)$, and if we let $m_H(t)=m_H(N_t)$ then \begin{equation} \label{WeakGerochMonotonicity}
    \frac{dm_H}{dt}(t) = \sqrt{ \frac{|N_t|}{16 \pi} } \left[ \frac{1}{2}  + \frac{1}{16\pi} \int_{N_t} \left( 2\frac{|\nabla_{N_t} H_{N_t}|^2}{H_{N_t}^2} + R -2K_{N_t} +\frac{1}{2} (\lambda_1 - \lambda_2)^2    \right) dN_t       \right]
\end{equation} 
for almost every $t$, where $K$ is the Gaussian curvature and $\lambda_i$ are the principal curvatures of the flow surfaces. Usually, the way that this formula is used is that assuming $R\geq 0$ and applying the weak Gauss-Bonnet formula, one concludes that the Hawking mass is monotonic under the (weak) IMCF. Starting the flow from an outermost minimal surface, one then obtains the Riemannian Penrose inequality. However, the formula continues to hold even if $R \ngeq 0$, and this will be useful later.

All of these results are found in \cite{HuiskenIlmanen}.
Finally, it is well known that if $S_t$ is a flow of surfaces going off to infinity, where each surface is homotopic to the 2-sphere, then
\begin{equation*}
    \lim_{t \rightarrow \infty} m_H(S_t) = E_{ADM}.
\end{equation*}

\subsection{Spherically Symmetric Metrics} Here we collect some facts about spherically symmetric metrics. We will work in spherical coordinates and our manifold will be
\begin{equation} \label{Coordinates}
    M=\mathbb{R}^3 \setminus B_\rho(0)=\lbrace (r, \theta, \phi) \;  | \;  r\in [\rho, \infty), \; \theta\in (0, \pi), \; \phi\in [0, 2\pi)  \rbrace.
\end{equation}
with metric
\begin{equation} \label{g}
    g=h(r)dr^2+r^2 d\Omega^2 = h(r) dr^2 + r^2(d\theta^2 + \sin^2(\theta) d\phi^2).
\end{equation}
Notice that in this metric, a coodrinate sphere $S_r$ has area $|S_r|=4\pi r^2$. A more general spherically symmetric metric has the form $h(r)dr^2+\rho^2(r) d\Omega^2$, however we will have no need of considering such metrics. We remark some similar calculations in the case of rotationally symmetric metrics were done by Lee and Sormani in \cite{LeeSormani}.

For a spherically symmetric initial data set, the general extrinsic curvature for the metric \eqref{g} has the form
\begin{equation*}
    k=hk_a dr^2 + k_br^2 d\theta^2 + k_b r^2 \sin^2(\theta) d\phi^2
\end{equation*}
where $k_a=k_a(r)$ and $k_b=k_b(r)$ are arbitrary functions of only $r$ (see \cite{BrayKhuri}). A calculation then shows that 
\begin{equation*}
    Tr_{{g}}k=k_a + 2 k_b
\end{equation*}
and 
\begin{equation*}
    |k|^2=k_a^2+2k_b^2
\end{equation*}
and so 
\begin{equation*}
    16\pi {\mu}={R} + 4k_a k_b + 2k_b^2.
\end{equation*}

\begin{proposition}
The scalar curvature ${R}$ of the metric \eqref{g} is given by
\begin{equation} \label{Rbar}
    R(r)=\frac{2h'(r)}{rh^2(r)} - \frac{2}{r^2 h(r)} + \frac{2}{r^2}.
\end{equation}

\end{proposition}

\begin{proof}
This can be established by a direct calculation using the definition of scalar curvature.
\end{proof}

\begin{proposition} \label{MeanCurvatureFormula}
The mean curvature $H$ of a coordinate sphere $S_{r}$ having radius $r$ in the metric \eqref{g} is given by
\begin{equation} \label{H}
  H_{ S_{ r }   } = \frac{2}{ r   \sqrt{h( r  )}  }.
\end{equation}
\end{proposition}

\begin{proof}
Again this can be calculated directly.
\end{proof}

Oftentimes we will write
\begin{equation*} 
    R=\frac{2h'   }{rh^2   } - \frac{2}{r^2 h    } + \frac{2}{r^2}
\end{equation*}
and
\begin{equation*} 
  H_{ S_{ r }   } = \frac{2}{ r   \sqrt{h    }  }
\end{equation*}
for short. 

\begin{remark}
These formulas can also be found in any standard Riemannian geometry textbook, though usually there the metric is given in the form $g=ds^2 + r^2(s) d\Omega^2$. Making the substitution $ds=\sqrt{h}dr$ one can obtain the above formulas. See also equation 4.1 and the first equation at the top of page 11 of \cite{BrayKhuri}.
\end{remark}

\begin{remark}
We remark that normally one expects higher order derivatives to appear in the formula for the scalar curvature since it involves derivatives of the Christoffel symbols. However, in the case of spherical symmetry these higher derivatives cancel out. Hence the above expression makes sense even if the metric is only $C^1$. 
\end{remark}

\begin{proposition}
The Hawking mass for coordinate spheres $S_{r}$ in the metric \eqref{g} is given by
\begin{equation} \label{SphericalHM}
    m_H(S_{r})=\frac{r}{2} \left( 1 - \frac{1}{h(r)}   \right).    
\end{equation}
\end{proposition}

\begin{proof}
This follows from the definition of the Hawking mass and Proposition \ref{MeanCurvatureFormula}.
\end{proof}

\begin{proposition} \label{IMCFFlow}
Let $S_0$ be some particular initial coordinate sphere and $|S_0|=4\pi r_0^2$ its area. Then the coordinate sphere flow $S(t)=S_t$ defined by
\begin{equation}
    \left(r(t), \theta(t), \phi(t)\right)=\left(r_0 e^{t/2}, \theta(0), \phi(0)\right),  \quad t\in[0, \infty)
\end{equation}
is an inverse mean curvature flow. 
\end{proposition}

\begin{proof}
Calculating, we have
\begin{equation*}
    r'(t)=\frac{r_0}{2}e^{t/2} = \frac{1}{2}r(t)
\end{equation*}
and so the velocity squared of the flow is
\begin{equation*}
    v^2(t)={g}(r'(t) \partial_{r}, r'(t) \partial_{r}   ) = h(r(t)) \left(  \frac{r_0}{2}e^{t/2} \right)^2 = \frac{h(r(t))}{4} \left(  {r_0}e^{t/2} \right)^2 = \frac{h(r)r^2(t)}{4} = \frac{1}{H_{Sr}^2}
\end{equation*}
by Proposition \ref{MeanCurvatureFormula} as desired.
\end{proof}

\begin{remark}
This calculation also serves to double check our formula for the mean curvature since it is well known that surfaces evolving under the IMCF satisfy $|N_t|=A_0 e^t$ for some constant $A_0$, see \cite{HuiskenIlmanen}.
\end{remark}

\begin{proposition}
If we write $h(r)=1+\varphi(r)$, then the ADM energy $E_{ADM}$ of the metric \eqref{g} is given by 
\begin{equation*}
     E_{ADM}=\lim_{r \rightarrow \infty} \frac{r}{2}\frac{\varphi(r)}{\sqrt{h(r)}}. 
\end{equation*}
In addition, if $\lim_{r\rightarrow \infty} h(r) = 1$ then 
\begin{equation*}
    E_{ADM}= \lim_{r \rightarrow \infty} m_H(S_r).
\end{equation*}
\end{proposition}

\begin{proof}
The first equation follows from the definition of the ADM energy. For the second, we substitute $h=1+\varphi$ in \eqref{SphericalHM} and use the assumption that $h \rightarrow 1$ to obtain
\begin{equation}
    E_{ADM}=\lim_{r \rightarrow \infty} \frac{r}{2}\frac{\varphi(r)}{\sqrt{h(r)}} =  \lim_{r \rightarrow \infty} \frac{r}{2}\varphi(r) = \lim_{r \rightarrow \infty} \frac{r}{2}\frac{\varphi(r)}{h(r)} = \lim_{r \rightarrow \infty} M_H(S_{r})
\end{equation}
\end{proof}

\begin{remark}
Again, this calculation is not strictly necessary since it is well known that in this case the Hawking mass will converge to the ADM energy, see \cite{Wu} for example, but once more it serves as a check of the spherically symmetric formulas.
\end{remark}

Using the Geroch monotonicity formula \eqref{WeakGerochMonotonicity} applied to the flow of Proposition \ref{IMCFFlow} we see that since the flow surfaces are spheres all the terms except the scalar curvature disappear (we use the Gauss-Bonnet formula to get $\int_{S_t} K_{S_t} dS_t =2\pi \chi(S_t)=4\pi$) and obtain
\begin{equation*}
    \frac{d m_H}{dt}(t)=\sqrt{\frac{|S_t|}{16\pi}}\left[ \frac{1}{16\pi} \int_{S_t} R \, dS_t  \right]
\end{equation*}
and so it is not necessary to check this explicitly in spherical symmetry. However, we can do this calculation explicitly to double check our formulas. In particular, the formula \eqref{Rbar} is the most important formula in the paper and so we want to again check that it is correct. As a result, we have the following proposition which serves as a check.

\begin{proposition} \label{MonotonicitySpherical}
Let $S(t)$ be the flow of Proposition \ref{IMCFFlow} and let
\begin{equation}\label{SPhericalHM2}
    m_H(t)=m_H(S(t))=\frac{r(t)}{2}\left(  1 - \frac{1}{h(r(t))}   \right)
\end{equation}
be the Hawking mass of the flow surfaces. Then
\begin{equation*} 
    \frac{d m_H}{dt}(t)=\sqrt{\frac{|S_t|}{16\pi}}\left[ \frac{1}{16\pi} \int_{S_t} {R} \, dS_t  \right].
    \end{equation*}
\end{proposition}

\begin{proof}
Differentiating we obtain
\begin{align*}
    \frac{d m}{dt}(t)&=\frac{r'(t)}{2}\left(1-\frac{1}{h(r(t))}\right)+\frac{r(t)}{2}\left(  \frac{h'(r(t)) r'(t) }{h^2(r(t))}    \right) \\
    &=\frac{r(t)}{4}\left(1-\frac{1}{h(r(t))}\right)+\frac{r(t)}{4}\left(  \frac{h'(r(t)) r(t) }{h^2(r(t))}    \right)\\
    &=\frac{r(t)}{4}\left(  1-\frac{1}{h(r(t))} + \frac{r(t)h'(r(t))}{h^2(r(t))}  \right)
\end{align*}
where we used $r'=r/2$. On the other hand
\begin{align*}
    \sqrt{\frac{|S_t|}{16\pi}}\left[ \frac{1}{16\pi} \int_{S_t} {R} \, dS_t  \right]&= \frac{r(t)}{2} \left[ \frac{4\pi r^2(t)}{16\pi} \left( \frac{2h'(r(t))   }{r(t)h^2(r(t))   } - \frac{2}{r^2(t) h(r(t))    } + \frac{2}{r^2(t)}     \right)    \right] \\
     &=\frac{r(t)}{4}\left(  1-\frac{1}{h(r(t))} + \frac{r(t)h'(r(t))}{h^2(r(t))}  \right)
\end{align*}
and so the two sides are equal as desired.
\end{proof}

\section{Proof of Theorem \ref{MainTheorem}} \label{Proof}

The basic idea will be to prescribe negative scalar curvature on our spherically symmetric manifold by solving equation \eqref{Rbar}, defining an appropriate $k$ so that the weak energy condition is satisfied, the boundary is an apparent horizon, the data set is asymptotically flat, and checking that the conclusions of Theorem \ref{MainTheorem} hold.

For simplicity, we let $\rho=1$ so that our manifold will be 
\begin{equation*}
    M=\mathbb{R}^3 \setminus B_1 (0)=\lbrace (r, \theta, \phi) \;  | \;  r\in [1, \infty), \; \theta\in (0, \pi), \; \phi\in [0, 2\pi)  \rbrace.
\end{equation*}
Next, for $n>1$ we define a smooth cut-off function $0 \leq \Phi_n(r)\in C^{\infty}([0, \infty))$ which satisfies
\begin{align}
    \Phi_n(r) \coloneqq \begin{dcases}
    1 &\quad \text{if} \quad |r|\leq n \\
    \text{smooth, decreasing} &\quad \text{if} \quad n<r < n+1 \\
    0 &\quad \text{if} \quad r\geq n+1
    \end{dcases}.
\end{align}
Next, we consider an IMCF starting from the coordinate sphere $r_0=1$. Then the flow is given by 
\begin{equation*}
    r(t)=e^{t/2}, \quad t\in [0, \infty)
\end{equation*}
and using the Geroch monotonicity in spherical symmetry we obtain
\begin{equation*} 
    \frac{d m_H}{dt}(t)=\sqrt{\frac{|S_t|}{16\pi}}\left[ \frac{1}{16\pi} \int_{S_t} R \, dS_t  \right] = \frac{r(t)}{2} \left( \frac{1}{4}R(r(t)) r^2(t)  \right) = \frac{1}{8} R(r(t))r^3(t) = \frac{1}{8} R(e^{t/2}) e^{3t/2}
    \end{equation*}
and so
\begin{equation}\label{MH1}
    m_H(\infty)-m_H(0)=\int_0^\infty \frac{1}{8}R(e^{t/2}) e^{3t/2} dt.
\end{equation}
We will take 
\begin{equation*}
    R_n(r)=\mathcal{R}_n \Phi^2_n(r)
\end{equation*}
where we define the constant $\mathcal{R}_n$ by the condition
\begin{equation}\label{MH2}
    \int_0^\infty \frac{1}{8}R_n(e^{t/2}) e^{3t/2} dt= \int_0^\infty \frac{1}{8}\mathcal{R}_n \Phi^2_n(e^{t/2}) e^{3t/2} dt = -\frac{3}{4}
\end{equation}
and then we easily see that
\begin{equation*}
    \mathcal{R}_n < 0, \quad \lim_{n \rightarrow \infty} \mathcal{R}_n = 0.
\end{equation*}
We will take $R_n(r)$ to be our prescribed scalar curvature. Before we construct the metric however, we construct the extrinsic curvature we need to satisfy the weak energy condition. A calculation then shows that 
\begin{equation*}
    Tr_{{g}}k=k_a + 2 k_b
\end{equation*}
and 
\begin{equation*}
    |k|^2=k_a^2+2k_b^2
\end{equation*}
and so 
\begin{equation*}
    16\pi {\mu}=R+(Tr_g k)^2 -|k|^2={R} + 4k_a k_b + 2k_b^2.
\end{equation*}
Thus, the simplest choice is to let 
\begin{equation*}
    k_a(r)=0, \quad k_b^{\mp}(r)= \pm\sqrt{\frac{|\mathcal{R}_n|}{2}} \Phi_n(r)
\end{equation*}
and define
\begin{equation} \label{DefinitionOfk}
    k_n^\mp(r)= \pm \sqrt{\frac{|\mathcal{R}_n|}{2}} \Phi_n(r) \left(r^2 d\theta^2 + r^2 \sin^2(\theta) d\phi^2 \right).
\end{equation}
Whether we choose the $+$ or $-$ will depend on whether we want the boundary to be a future or past apparent horizon. We will want it to be a future apparent horizon.

With this $k_n$ and $R_n$ we get
\begin{equation*}
    16\pi \mu_n = R_n+(Tr_g k_n)^2 -|k_n|^2={R}_n + 2k_b^2 = \mathcal{R}_n \Phi^2_n(r) + \left\vert\mathcal{R}_n \right\vert \Phi^2_n(r) = 0
\end{equation*}
since $\mathcal{R}_n<0$. Thus for these choices of scalar and extrinsic curvatures, the weak energy condition is satisfied. Notice, both $R_n$ and $k_n$ are compactly supported. 

The next step is to construct a spherically symmetric metric with scalar curvature $R_n(r)$. To do this, we plug it into \eqref{Rbar} for $R$ and analyze the resulting ordinary differential equation. Rearranging, we obtain
\begin{equation*}
    h'=\frac{h}{r} - \frac{h^2}{r} + \frac{1}{2}R_n rh^2
\end{equation*}
and plugging in $R_n=\mathcal{R}_n \Phi_n^2 = -|\mathcal{R}_n| \Phi_n^2$ we obtain
\begin{equation} \label{ODE}
    h'=\frac{h}{r} - \left(  \frac{1}{r} + \frac{1}{2} r|\mathcal{R}_n| \Phi_n^2    \right)h^2
\end{equation}
and establishing the existence of a solution $h(r)$ for $r\in[1, \infty)$ for any initial condition $h(1)>0$ turns out to be quite easy.

\begin{proposition} \label{Bounds}
For any $h(1)>0$ and any $n>1$ there exist constants $L_n$ and $U_n$ such that the solution of \eqref{ODE} satisfies
\begin{equation}
    0<L_n < h(r) < U_n
\end{equation}
for all $r\in [1, r^*)$ where $[1, r^*)$ is the maximal interval of existence for the solution.
\end{proposition}

\begin{proof}
First, we claim there is a constant $C_n>0$ such that 
\begin{equation*}
    \frac{1}{r} + \frac{1}{2} r|\mathcal{R}_n| \Phi_n^2 < \frac{C_n}{r}
\end{equation*}
for $r\in[1, \infty)$. To see this, notice that $\frac{1}{2} r|\mathcal{R}_n| \Phi_n^2=0$ for $r\geq n+1$ and $\frac{1}{2} r|\mathcal{R}_n| \Phi_n^2 \leq \frac{1}{2}(n+1)|\mathcal{R}_n| \coloneqq B_n$ for $1\leq r \leq n+1$. Thus, 
\begin{equation*}
    \frac{1}{2} r|\mathcal{R}_n| \Phi_n^2 \leq \frac{(n+1)B_n}{r}
\end{equation*}
and so 
\begin{equation*}
    \frac{1}{r} + \frac{1}{2} r|\mathcal{R}_n| \Phi_n^2 \leq \frac{1}{r} + \frac{(n+1)B_n}{r}
\end{equation*}
and so $C_n\coloneqq 2+(n+1)B_n$ does the trick. Also notice $1/C_n<1/2$ for all $n$.

Now choose $L_n$ to be any number such that 
\begin{equation} \label{LowerBound}
    0<L_n<\min \left\lbrace h(1), \frac{1}{C_n} \right\rbrace.
\end{equation}
We claim that this number acts as a barrier for the solution. For, suppose that $s>1$ is the first value of $r$ at which $h(s)=L_n$. since the solution starts out larger than $L_n$, we must have $h'(s)\leq 0$. But at $s$ we have
\begin{equation*}
    h'(s)=\frac{L_n}{s}-\left( \frac{1}{s} + \frac{1}{2}s|\mathcal{R}_n|\Phi^2_n(s)   \right) L_n^2 > \frac{L_n}{s}-\frac{C_n}{s} L_n^2 > \frac{L_n}{s} - \frac{L_n}{s}=0
\end{equation*}
yielding a contradiction. Hence, there is no such smallest $s$ and thus $h(r)>L_n$ on $[1, r^*)$.

Similarly, we can show the existence of the upper bound. Let $U_n$ be any number such that 
\begin{equation} \label{UpperBound}
    U_n > \max \left\lbrace h(1), 1 \right\rbrace
\end{equation}
and again let $s$ be the smallest value of $r$ where $h(s)=U_n$. Since the solution starts out smaller, we must have $h'(s)\geq 0$. But we have
\begin{equation*}
    h'(s)=\frac{U_n}{s} - \left(  \frac{1}{s} + \frac{1}{2} s|\mathcal{R}_n| \Phi_n^2(s)    \right)U_n^2 \leq \frac{U_n}{s}-\frac{U_n^2}{s}<0
\end{equation*}
again yielding a contradiction. Hence, there is no such smallest $s$ and $h(r)<U_n$ for all $r\in [1, r^*)$. \end{proof}

\begin{proposition} \label{Existence}
For any $h(1)>0$ and any $n>1$, the differential equation \eqref{ODE} possesses a unique smooth solution $h(r)>0$ for $r\in [1, \infty)$.
\end{proposition}

\begin{proof}
The existence, uniqueness, and lower bound of the solution for all $r\in [1, \infty)$ follows easily from the apriori bounds of Proposition \ref{Bounds} and the Picard-Lindel\"{o}f theorem. The smoothness follows from the smoothness of $\Phi_n(r)$.
\end{proof}

\begin{proposition} \label{AsymptoticBehavior}
For each $k\geq 0$ there exists a constant $C_k$ such that solution given in Proposition \ref{Existence} satisfies
\begin{equation} \label{FallOffh}
    \left\vert \frac{d^k}{dr^k} \left( h(r)-1  \right) \right\vert \leq \frac{C_k}{r^{1+k}}
\end{equation}
for $r\in [1, \infty)$.
\end{proposition}

\begin{proof}
Consider the smooth unique solution of Proposition \ref{Existence}. Since it exists for all $r\in [1, \infty)$ it has some value at $r=n+1$, call it $\mathcal{A}=h(n+1)>0$ which depends on $h(1)$. Now, for $r\geq n+1$ the equation \eqref{ODE}  simplifies to
\begin{equation} \label{ODE2}
    \mathcal{H}'=\frac{\mathcal{H}}{r}-\frac{\mathcal{H}^2}{r}
\end{equation}
since $\Phi_n(r)=0$ for $r\geq n+1$. By uniqueness of $h(r)$, we can solve \eqref{ODE2} with the initial condition $\mathcal{H}(n+1)=\mathcal{A}=h(n+1)$ and the resulting solution $\mathcal{H}(r)$ will coincide with the solution $h(r)$ given by Proposition \ref{Existence} on $[n+1, \infty)$. Fortunately, \eqref{ODE2} is a separable differential equation, and we can calculate the solution explicitly which is 
\begin{equation*}
    \mathcal{H}(r)=\frac{r}{C+r}
\end{equation*}
where to impose our initial condition we must take
\begin{equation*}
    C=\frac{(n+1)(1-\mathcal{A})}{\mathcal{A}}.
\end{equation*}
Notice that with this $C$ the solution does indeed exist on all of $[n+1, \infty)$. If $\mathcal{A}=1$ then $\mathcal{H}(r)=1$. If $0<\mathcal{A}<1$ then $C>0$ and so the singularity would occur at some $r<0$ which is outside of our interval of interest. If $\mathcal{A}>1$ then the singularity would occur at
\begin{equation*}
    r=-C=\frac{(n+1)(\mathcal{A}-1)}{\mathcal{A}}<n+1
\end{equation*}
which again is outside our interval of interest.

Now we can write the solution as 
\begin{equation*}
    \mathcal{H}(r)=1-\frac{C}{C+r}
\end{equation*}
from which \eqref{FallOffh} follows. 
\end{proof}

The bound \eqref{FallOffh} implies that the resulting metric given by \eqref{g} is asymptotically flat. Now we want $S_1(0)=S_1=\partial M$ to be a future apparent horizon so that
\begin{equation*}
    \theta_+(S_1) = H_{S_1} + Tr_{S_1}k=0.
\end{equation*}
Since we have 
\begin{equation*}
    H_{S_1}=\frac{2}{\sqrt{h(1)}}
\end{equation*}
by \eqref{H}, we need $k_b(1)$ to be negative, so we take
\begin{equation*}
    k(r)=k_n^+(r)=-\sqrt{\frac{|\mathcal{R}_n|}{2}} \Phi_n(r) \left(r^2 d\theta^2 + r^2 \sin^2(\theta) d\phi^2 \right)
\end{equation*}
and so 
\begin{equation*}
    Tr_{S_1}k =2k_b = -2 \sqrt{\frac{|\mathcal{R}_n|}{2}} \Phi_n(1) = -2\sqrt{\frac{|\mathcal{R}_n|}{2}}
    \end{equation*}
Now, to obtain $\theta_+=0$ we must make the choice
\begin{equation*}
    h(1)=\frac{2}{|\mathcal{R}_n|}
\end{equation*}
which we use as the initial condition for \eqref{ODE}. Notice since as $n\rightarrow \infty$ we have $\mathcal{R}_n \rightarrow 0$, then $h(1)\rightarrow \infty$ and $H_{S_1}\rightarrow 0$. 

Now, $S_1=\partial M$ is a future apparent horizon, but it is not necessarily outermost. However, by Theorem 1.3 in \cite{AnderssonMetzger}, there exists a unique outermost future apparent horizon, which we denote by $\tilde{S}$ (hich depends on $n$). We remark that, as stated by the authors, the results in that paper do not depend on the choice of energy conditions in anyway. By spherical symmetry and uniqueness, $\tilde{S}$ must by a sphere of some radius $\tilde{r}\geq 1$ and since our metric is of the form \eqref{g} we have $\tilde{A}=4\pi \tilde{r}^2 \geq 4\pi= |S_1|$. Now we take the outermost minimal area enclosure of $\tilde{S}$, denoted by $\Sigma(\tilde{S})$ which exists and is unique by the results in \cite{BassaneziTamanini}, see the comment after Definition 10 in \cite{Bray}.  

Moreover, due to the spherical symmetry of $\tilde{S}$ and the uniqueness the minimal area enclosure of $\tilde{S}$ must also be spherically symmetric, and since spheres of larger $r$ have larger area by the form of the metric, we have $\Sigma{\tilde{S}}=\tilde{S}$ and so
\begin{equation*}
    \sqrt{  \frac{A}{16\pi}  } = \sqrt{  \frac{A_{min}(\tilde{S})}{16\pi}  } = \sqrt{  \frac{|\tilde{S}|}{16\pi}  } \geq \sqrt{  \frac{|S_1|}{16\pi}  } = \frac{1}{2}.
\end{equation*}

Now we also have
\begin{equation*}
    E_{ADM}=m_{H}(\infty)=m_{H}(S_1)+ (m_{H}(\infty)- m_{H}(S_1) )= m_{H}(S_1) - \frac{3}{4}
\end{equation*}
by \eqref{MH1} and \eqref{MH2}. But
\begin{equation*}
    m_H(S_1)=\frac{1}{2} \left( 1-\frac{1}{h(1)}   \right)
\end{equation*}
by \eqref{SphericalHM} and since $h(1)\rightarrow \infty$ as $n\rightarrow \infty$ we have $E_{ADM}\rightarrow -1/4$ and so for sufficiently large $n$ we have  
\begin{equation*}
    \sqrt{\frac{{A}}{16\pi}} \geq \frac{1}{2} >0 > E_{ADM}
\end{equation*}
and thus for sufficiently large $n$ we can take the exterior of $\tilde{S}$ with the constructed $g$ and $k$ to be the counter example satisfying \eqref{MainTheoremInequality1}.

Since $k$ is compactly supported by \eqref{ADMMomentum} we have that $P_i=0$ and so
\begin{equation*}
    m_{ADM}=|E_{ADM}|
\end{equation*}
by \eqref{ADMMass}. But then we have 
\begin{equation*}
    \lim_{n\rightarrow \infty} m_{ADM} =\frac{1}{4}
\end{equation*}
and so for sufficiently large $n$ we obtain
\begin{equation*}
    \sqrt{\frac{{A}}{16\pi}} \geq \frac{1}{2} >m_{ADM}
\end{equation*}
yielding the counterexample to \eqref{MainTheoremInequality2}, completing the proof.

Notice, as a byproduct we constructed an asymptotically flat initial data set satisfying the weak energy condition with negative ADM energy. However, our example doesn't technically satisfy all of the assumptions of the Schoen and Yau theorem with the exception of the energy condition, since it has a boundary. However, it is quite easy to make some minor changes to the proof to have no boundary, as we do in the next section. 

\section{Proof of Theorem \ref{MainTheorem2}}

The idea of the proof here is almost the same as for Theorem \eqref{MainTheorem}, except we prescribe a slightly different choice scalar curvature. We take
\begin{equation*}
    M=\mathbb{R}^3 =\lbrace (r, \theta, \phi) \;  | \;  r\in [0, \infty), \; \theta\in (0, \pi), \; \phi\in [0, 2\pi)  \rbrace.
\end{equation*}
We take any function $\eta(r) \in C^{\infty}_c(\mathbb{R})$ with the properties 
\begin{equation*}
    0\leq \eta(r) \leq 1, \quad \text{spt}({\eta}) \subset (1, 2) 
\end{equation*}
so that it vanishes outside of $(1, 2)$.

Next, for our prescribed scalar curvature we will take
\begin{equation*}
    R_{\varepsilon}(r)=-\varepsilon \eta^2(r)
\end{equation*}
where $\varepsilon>0$ is some constant. 
Thus, we seek to solve

\begin{equation*}
    h'=\frac{h}{r} - \frac{h^2}{r} + \frac{1}{2}R_{\varepsilon}rh^2
\end{equation*}
which upon rearranging we obtain
\begin{equation} \label{ODEPM}
    h'=\frac{h}{r} - \left(  \frac{1}{r} + \frac{1}{2} r\varepsilon \eta^2    \right)h^2
\end{equation}
and we wish to solve this on $[0, \infty)$. Now, it looks like this differential equation has a singularity at $r=0$ and there might be some problems obtaining smooth solutions. However, with the correct choice of initial condition, any such problems disappear. 

We take as our initial condition $h(0)=1$. Notice, that since $\eta$ is only supported for $1<r<2$, then on $[0, 1]$ the differential equation simplifies to \begin{equation*}
    h'=\frac{h}{r} - \frac{h^2}{r}
\end{equation*}
and for $h(0)=1$ we have that $h(r)\equiv 1$ is a smooth solution. Moreover, in that case the metric for $0\leq r<1$ is then
\begin{equation*}
    dr^2 + r^2(d\theta^2 + \sin^2(\theta) d\phi^2) 
\end{equation*}
which is just the Euclidean metric in spherical coordinates. Thus, we will have $g=\delta$ for $r<1$. Next, for $r\geq 1/2$ we solve \eqref{ODEPM} with the initial condition $h(1/2)=1$. Propositions \ref{Bounds}, \ref{Existence}, and \ref{AsymptoticBehavior} all apply with the same proofs and the resulting solution $h(r)>0$ is smooth on $[1/2, \infty)$ and identically $1$ on $[1/2, 1]$ by uniqueness, and thus we obtain a smooth solution on all of $[0, \infty)$ with the appropriate asymptotic behavior. 

To obtain the weak energy condition, we define
\begin{equation*}
    k_a(r)=0, \quad k_b(r)=\sqrt{\frac{\varepsilon}{2}} \eta(r)
\end{equation*}
so that
\begin{equation*}
    k=\sqrt{\frac{\varepsilon}{2}} \eta(r) \left( r^2 d\theta^2 + r^2 \sin^2(\theta) d\phi^2  \right)
\end{equation*}
as we did before, taking the square root.

Next, we look at the ADM energy of this metric. We see $M_H(S_1)=0$ since $h(r)=1$ for $0\leq r\leq 1$ by \eqref{SphericalHM}. Once again using the IMCF starting from $S_1$ given by $r(t)=e^{t/2}$ we obtain
\begin{align*}
    E_{ADM}=m_{H}(\infty)=m_{H}(S_1)+ (m_{H}(\infty)- m_{H}(S_1) )&=  \int_0^\infty \frac{1}{8}R_\varepsilon(e^{t/2}) e^{3t/2} dt \\
    &= -\int_0^\infty \frac{1}{8} \varepsilon \eta^2_n(e^{t/2}) e^{3t/2} dt <0
\end{align*}
and so the positive mass theorem is violated. 

Moreover, the resulting data set $(\mathbb{R}^3, g, k)$ does not contain any minimal surfaces. For suppose it did. Then by the results of Meeks, Simon, and Yau \cite{MeeksSimonYau} there would be a unique outermost minimal surface with possibly multiple components. By spherical symmetry, this outermost minimal surface would have to be invariant under all rotations, and hence it would have to have one component and be a sphere of some radius $r$. But, since $h(r)$ is smooth and bounded, we see by \eqref{H} that $H_{S_r}>0$ for any $r$, yielding a contradiction. Hence, there are no minimal surfaces in the data set.

Now, suppose $(\mathbb{R}^3, g, k)$ contains a future apparent horizon, that is a surface with $\theta_+=0$. Again, by Theorem 1.3 in \cite{AnderssonMetzger} there would be a unique outermost future apparent horizon which by uniqueness and spherical symmetry of the data set would have to be a sphere of some radius $r$. But for such a sphere we have
\begin{equation*}
    \theta_+(S_r) = H_{S_r} + 2k_b(r) >0
\end{equation*}
by our definition of $k$ yielding a contradiction. Hence there are no future apparent horizons, completing the proof.

\section{Conclusion}

We see that the various geometric inequalities in general relativity depend very strongly on the choice of energy condition assumed by the matter fields. Of course, with the dominant energy condition being stronger than the weak energy condition, there is still a (high) chance that the Penrose conjecture, in either of the forms we have stated holds in the case of the dominant energy condition, as evidenced by it holding under this condition in the case of spherical symmetry \cite{Hayward}, and with the interesting very plausible general approach in \cite{BrayKhuri}. Of course with the positive mass theorem holding in the case of the dominant energy condition \cite{SchoenYau2} provides further indirect evidence. 

Moreover, the dominant energy condition giving an upper bound for the speed at which mass-energy can flow in terms of the speed of light is physically more palatable than merely the weak energy condition. Nevertheless, since Penrose's heuristic argument seems to still hold with the weak energy condition, it is important to pinpoint exactly where the problem is. It might be that there is some subtle assumption where an upper bound for the speed of mass-energy might be useful. It might lie in the assumption that the system eventually settles down to equilibrium. Some physical insight such as this might suggest a path to proving the Penrose conjecture in the case of the dominant energy condition.

\medskip

\noindent \textbf{Acknowledgements} I would like to thank Marcus Khuri for helpful discussions.

\bibliographystyle{model1-num-names}
\bibliography{ref_new}

 \footnotesize

  J.S.~Jaracz, \textsc{Department of Mathematics, Texas State University,
    San Marcos, TX 78666}\par\nopagebreak
  \textit{E-mail address} \texttt{jsj74@txstate.edu}

\end{document}